\let\ti=\times

\let\la=\lambda
\let\al=\alpha
\let\va=\varphi

\def\que#1#2{\displaystyle\frac{#1}{#2}}

\documentclass[pra,twocolumn]{revtex4}

\usepackage{amsfonts,amssymb,bm,graphicx}

\begin{document}

\title{Basic outlines of a new hypothesis on physical reality}

\author{R. Alvargonz\'alez and L. S. Soto}

\affiliation{Facultad de F\'{\i}sica,  Universidad Complutense,
Madrid, Spain}

\date{\today}

\begin{abstract}
The papers mentioned in the bibliography lead to this new hypothesis which
constitutes a wide panorama of the physical reality. Its coherence and its
simplicity are virtues that make interesant to gaze upon it.

We feel grateful to M. J. Sparnaay and to S. K. Lamoreaux whose measurements
on zero-point radiation have been a strong link between our reasonings and the
physical reality, and we also feel grateful to Timothy H. Boyer whose works on
zero-point radiation have been a welcomed help.
\end{abstract}

\maketitle

\section{Basic outlines at cosmologica level (see [1] and [2])}

\begin{enumerate}\def\labelenumi{\arabic{section}.\arabic{enumi})}\itemsep=1pt
\item The Universe began with the breaking in of photons of high energy,
contained within the 3-dimensional spherical surface
$w^2+x^2+y^2+z^2=(R_i)^2$.
\item Photons are ``packets of energy'' which are characterised by travelling
at the same velocity, $c$, and possessing the same quantity of action, $h$.
They are differentiated by their wavelengths, $\la$, and their energies,
$E_\la=hc/\la$.
\item The initial photons were distributed equally in every possible
direction. As a result, the Universe has continued to expand at the speed of
light in every direction, since its beginning at $t=0$. Therefore, when a
period of time $t$ has elapsed from $t=0$, the Universe is configured as the
3-dimensional spherical surface $w^2+x^2+y^2+z^2=(R_i+ct)^2$.
\item The condition $w^2+x^2+y^2+z^2=(R_i+ct)^2$ determines that every point
of space is subjected to a tension, whose intensity $E$ is directly
related to the curvature of space $1/(R_i+ct)$. The value of $E$ is
therefore the same at any moment, for all points in space, and lessens equally
at all of them with the passing of time.

\item The constant increase in the radius of the Universe implies a constant
reduction in $E$, and consequently, the existence of:
\def\labelitemi{--}
\begin{itemize}\itemsep=1pt
\item An energy flow, zero-point radiation, which is inherent to space, and
consequently possesses a relativistically invariant spectrum. For this to be
so, its photons must have trajectories distributed equally along all
directions of space, and the abundance of the photons must be inversely
proportional to the cubes of their wavelengths. The density of this energy
flow is directly proportional to the cube of the curvature.

\item A centrifugal energy flow, whose intensity is directly proportional to
the $4^{\rm th}$ power of the curvature.
\end{itemize}
\item The wavelength of the most energetic photons of zero-point radiation is directly
proportional to the length of the radius of the Universe, so that the density
of the energy of that radiation in the Universe lessens in direct proportion
to the cube of the radius, and the total amount of that energy remains
constant as the length of the radius increases.
\item Many of the initial photons mentioned in 1.2, become configured as
elementary particles. They revolve round axes passing through their centres,
and because of their configuration, do so in such a way as to possess a spin
of $\hbar/2$, and a tangential velocity of $c$ at all their points.
\item The velocity of the photons, $c$, is an absolute ceiling which cannot be attained
by elementary particles [6].
\item Elementary particles can come together to form higher-order entities,
atoms, which possess properties which go beyond those of their
components. Atoms can come together in higher-order entities, molecules, which
also possess properties which go beyond those of their components.
\item The objects which we can perceive are made up to aggregations of very
great numbers of atoms and molecules. The stars are enormous cosmic objects
which emit light.
\item The sun is a star, around which orbit many cosmic objects which are
unable to emit light. The planets, including Earth, are the 9 largest of these.
\item The galaxies are inmense structures, made up to thousands of millions of
stars. The sun is a star in the Milky Way, our galaxy. There are thousands of
millions of other galaxies at distances of millions of light-years from ours.
\item Within the Universe, whose radius is $R_U=R_i+ct$, we must distinguish
between the Material Universe, whose radius is $R_{U_m}<R_U$, and within it,
the Visible Universe, made up of those objects which we are able to observe or
which we could become able to observe, since the velocity at which they are
moving away, because of the expansion of the Material Universe, is less than
that of light.

The Visible Universe is therefore configured as a spherical dome on the
3-dimensional spherical surface $w^2+x^2+y^2+z^2=(R_{U_m})^2$, whose volume
is: $V_{U_v}=2\pi R_{U_m}(\va-\sin\va\cos\va)$, where $\va$ is the angle
between the radius which reaches the centre of the dome, and any of the radii
which reach the lesser circumference which forms its base. The volume of the
Visible Universe must therefore represent a fraction of the volume of the
Material Universe of $\que{\va-\que12\sin^2\va}{2\pi}.$
\item The quasars are visible cosmic objects whose light reaches us from the
area of the horizon of visibility, and their luminosity must have been that of
complete galaxies, i.e. a luminosity of hundreds of thousands of millions of
stars.

No quasars have been observed except in that area, which means that they are
structures belonging to the beginnings of the Universe, and ceased to exist
thousands of millions of years ago.
\item The configuration of the Universe as a 3-dimensional spherical surface
implies that space is relative. On a 2-dimensional spherical surface, we could
not distinguish any point by preference; its centre is
outside it. The same is true for a 3-dimensional spherical surface.

The evidence for the ``Big-Bang'', and the disappearance of the quasars implies
that time is not relative, but that it had a beginning at $t=0$ and $R_U=R_i$.
The configuration of the Universe as a spherical surface of radius
$R_U=R_i+ct$ implies that all its points lie at the same distance from its
centre, and that the value of $t$ is the same for all of them.

According to the Special Theory of Relativity, the laws of Physics are the
same at all points of the Material Universe, which requires that the lapse of
time between $t=0$ and $t_m$, the moment of generation of the last elementary
particle from the initial photons, is relatively in\-sig\-nificant.
\end{enumerate}

\section{Basic outlines at quantic level (see [3], [4] and [6]}

\begin{enumerate}\def\labelenumi{\arabic{section}.\arabic{enumi})}\itemsep=1pt
\item This paper uses the system of units $(e,m_e,c)$, in which the basic
magnitudes are the quantum of electrical charge, $e$, the mass of the
electron, $m_e$, and the speed of light, $c$. In this system, the unit of
length is $l_e=e^2m_e^{-1}c^{-2}$, and the unit of time is
$t_e=e^2m_e^{-1}c^{-3}$.
\item Photons, which are energy packets moving at the speed of light, $c$,
following rectilinear trajectories and possessing the same quantity of action,
$h=(2\pi/\al)\que{e^2}c=(2\pi/\al)m_el_e^2t_e^{-1}$, are differentiated by
their wavelengths $\la l_e$, and their energies
$E_{\la x}=hc/\la_x=(2\pi/\al)m_ec^2\cdot\la_x^{-1}$.
\item The substance of the photons undulates along cylindrical helices of
length $\la_x$ and radius $R=\la_x/\al$, which gives them a spin of $s=\hbar$.
The wavelength of the photon whose energy is $E_e=m_ec^2$ is
$\la_e=(2\pi/\al)l_e$, and that of the photon of energy $E_x=m_xc^2$ is
$\la_x=(2\pi/\al)\que{l_em_e}{m_x}$.
\item The transformation of a photon into an elementary particle does not
require any change in its substance. It remains the same energy packet, with a
quantity of action $h$ and wavelength $\la$, which, instead of moving in a
straight line as part of a series of waves, turns round on itself around a
sphere of radius $\al\la/2\pi$, thus becoming an separate entity, revolving
with a spin of $\hbar/2$ and tangential velocity $c$ at all its points.
\item The basic characteristics of elementary particles are:
\def\labelitemi{--}
\begin{itemize}\itemsep=1pt
\item Their mass, $m_x=E_x\cdot c^{-2}$, where $E_x=h_c\la_x^{-1}$ is the
energy of the photon of wavelength $\la_x$.
\item Their radius, $r_x=\al\la_x/2\pi$, where $\la_x$ is the wavelength of
the photon whose energy is $m_xc^2$.
\item Their spin $\hbar/2$, which derives from rotation around an axis passing
through the centre of the particle.
\item The equation $m_xr_x=e^2c^{-2}$, which is the primordial quantic
thres\-hold as concerns elementary particles.
\end{itemize}
\item The centrifugal force inherent to the spin of elementary particles with
a spin of $\hbar/2$ gives rise to the electrostatic force.
\item The electron is that elementary particle in which the centrifugal force
inherent to its spin is equal to the centripetal force generated by its
interactions with zero-point radiation.
\item The equation $m_xr_x=e^2c^{-2}$ requires that there cannot exist
elementary particles with a spin of $\hbar/2$ and a mass less than that of the
electron, nor with a radius greater than that of the electron, $r_e=1l_e$.
\item The centrifugal force of the particles of mass $m_x>m_e$ is equal to
$$\que{m_x(m_e)}{r_x(l_e)}\cdot c=m_x^2\que{m_el_e}{t_e^2}.$$
In other words, it is equal to that of the electron multiplied by
$(m_x/m_e)^2$. The surface of such a particle is equal to $4\pi(r_x)^2$, i. e.
$(m_e/m_x)^2$ times that of the electron, and therefore the centripetal force
which derives from its interactions with zero-point radiation is
$(m_e/m_x)^{-4}$ times that of the said centrifugal force.

In the case of the proton, $m_x=1836m_e$, and the centripetal force on the
surface of the particle is $1.1363\ti10^{13}$ times less than the
centrifugal force. The enormous imbalance between the centrifugal and
centripetal forces causes the appearance of the energy flows required to
balance them. These energy flows possess, at a distance of $1l_e$ from the
centre of the particle, an intensity per $l_e^2$ of $m_xl_e^2t_e^{-2}$ per
$t_e$, and at distance of $1l_x=l_em_e(m_x)^{-1}$, they have an intensity of
$m_x^2l_e^2t_e^{-2}$ per $l_e^2$ per $t_e$. In other words, they decrease in
proportion to the distance to the centre, and not in proportion to the square
of that distance. This is due to the interferences inherent to the small size
of the angles between the flows which fall on adjacent points, and to the
shortness of the distance between those points.

At distances from the centre which are greater than $1l_e$, the intensity of
these flows decreases according to their squares.
\item Keeping in mind what we have explained in 2.7, we seee that the energy
flows aroused in order to balance out the centripetal and centrifugal forces
on the surface of particles of mass $m_x>m_e$ cause an apparent attraction
between two such particles, which is equal to the gravitational attraction
between them. In the case of electrons, where there is no reason for such
energy flows to arise, the apparent attraction derives from the interactions
of the particle with zero-point radiation [6].
\item Again remembering what we have explained in 2.7, we realise that the
forces of repulsion inherent to the mutual electrostatic rejection between
protons, and the energy flows caused by the imbalance between the centripetal
and the centrifugal forces at the surfaces of
protons and neutrons, balance each other out within the atomic
nucleus, when the number of protons is approximately equal to the number of
neutrons, and the masses of both protons and neutrons are very approximately
equal to $1851m_e$. This explains the cohesion of atomic nuclei, the
characteristics of protons, and the strong interaction [6].

\item Neutrinos are particles which posses a mass very much smaller than that
of electrons, and have spin 1/2 and no charge.
\end{enumerate}

\section{Basic forces (see [3], [5] and [6]}
\begin{enumerate}\def\labelenumi{\arabic{section}.\arabic{enumi})}\itemsep=1pt
\item The interaction of zero-point radiation with elementary particles can
give rise to:
\def\labelitemi{--}
\begin{itemize}\itemsep=1pt
\item Forces of apparent attraction between any two particles, which are
directly proportional to the products of their masses
$$\left(m_x=\que{\al hc}{2\pi r_x}\cdot\que1{c^2}\right),$$
and inversely proportional to the squares of the distance between them:
$$f_G=\que{m_xm_y}{(d_{xy})^2}\cdot G,$$
where $f_G=$ gravitational attraction, $d_{xy}=$ the distance between the
particle with mass $m_x$ and that with mass $m_y$, and $r_x=$ the radius of
the particle with mass $m_x$. In other words, gravitational forces.
\item Forces opposing the change in the state of movement of any elementary
particle, which derive from the lateral relativistic Doppler effect on its
interactions with zero-point radiation, and which are proportional to the
product of the mass of the particle multiplied by the intensity
of the change in its state of movement, i. e. by the
acceleration
$$f_x=m_x a = m_x \left(\que{\partial v}{\partial t}\right).$$
In other words, inertial forces.
\item Forces deriving from the interaction of the rotation of elementary
particles with a spin of $\hbar/2$, with zero-point radiation. In other words,
electrostatic forces.
\item Forces of weak interaction, deriving from the interaction of the
rotation of elementary particles with zero-point radiation (still to be
demonstrated).
\end{itemize}
\item Forces of interaction arise between nucleons within the atomic nucleus.
These derive from the energy flows which arise to balance out the centrifugal
and centripetal forces on the surface of the nucleons, whose radii are limited
by the quantic threshold $m_xr_x=e^2c^{-2}$.
\end{enumerate}

We must consider, by comparison with the case of the electron, the curvature
of the Universe which would be required to explain these forces; this could be
the same curvature that Einstein suggested would be produced by the presence
of masses within space.

\section{Basic constants at the cosmological level (see [1])}
\begin{enumerate}\def\labelenumi{\arabic{section}.\arabic{enumi})}\itemsep=1pt
\item The Hubble constant, which is the velocity in km/s per megaparsec of
distance from Earth, at which cosmic objects move away, because of the
expansion of the Universe.
$$\que{60\ {\rm km/s}}{{\rm megaparsec}}<H_U<\que{90\ {\rm km/s}}{{\rm
megaparsec}};$$
$$\hbox{1 megaparsec}\ =3.26\times 10^6\
\hbox{light-years.}$$
Improving the precision of our knowledge of the value of the Hubble constant
is one of the most important challenges of cosmology at the start of the XXI
century.
\item The age of the Universe, $t_U$. Assuming $H_U=\que{70\ {\rm km/s}}{10^6\
{\rm parsec}},$ we calculate that $1.397\times 10^{10}$ years must have
elapsed in order for there to exist luminous objects which move away from us
at speeds near that of light, because of the expansion of the Universe.

The lack of precision in our knowledge of the value of the Hubble constant is
accompanied by an analogical imprecision in our knowledge of $t_U$. Also, the
value we have obtained is only a lower limit for $t_U$, since there could
exist cosmic objects which have moved away from us at speeds greater than that
of light, because of the Universe's expansion.

On the other hand, the age of the Universe would be less than what we have
calculated here if it proved possible to see two images of a very distant 
cosmic object by looking in two diametrically opposed directions. See 
figures 2 in [2] and 3 in [2], as well as the related texts.

\item The relation between the length of the radius of the Universe,
$R_U=1.39\times10^{10}$ light-years $=3.819306\ti10^{61} q_\la$, and the wavelength
of the photon possessing most energy in zero-point radiation
$x=5.259601\times10^{27}q_\la$, $k_U=R_U/x=7.264351\ti10^{33}$.
\item The gravitational constant $G$, whose value in the $(e,m_e,c)$ system is
expressed as $G=G_e\left(\que{e}{m_e}\right)^2$, where
$G_e=2.399998\ti10^{-43}$. The value of $G$ remains constant, and the
variations in the numerical coefficient $G_e$ are those which are required in
order that $G$ does not change, as wavelength $x$ of the photon which
possesses the greatest energy in zero-point radiation increases, in proportion
to the increase in the radius of the Universe.

The expression $G=\que{(q_\la)^2c^2}{2\pi *}$ is invariant, as will be seen in
5.1, 5.2 and 5.4, since $q_\la$, $c$, and $*$ are invariant.
\end{enumerate}

\section{Basic constants at the quantic level (see [6])}
\begin{enumerate}\def\labelenumi{\arabic{section}.\arabic{enumi})}\itemsep=1pt
\item The constant of the fine structure $\al$, which is the relation between
the wavelength $\la_E$ of the photon possessing an energy of $E=hc/\la_E$, and
the radius of the fermions of mass $E/c^2$.
\item The velocity of electromagnetic radiation through empty space, $c$. The
value of $c$ is an upper limit which cannot be attained by fermions.
\item The quantum of wavelength of electromagnetic radiation, $q_\la$, whose
measurement is $q_\la=(2\pi\al)^{1/2}L_P$, where
$L_P=\left(\que{\hbar G}{c^3}\right)^{1/2}$ is the Planck length.
\item The basic quantum $m_x  r_x=e^2 c^{-2}$, where $m_x$ is the mass of a
fermion and $r_x$ is its radius. This quantum, to which we will here give the
symbol $*$, is the necessary and sufficient condition for fermions to have a
spin of $\hbar/2$.
\item There exist quanta for other magnitudes, i. e. minimal quantities, such
that no smaller values could exist for them in physical reality, and that any
quantity of the said magnitude must be expressable in whole numbers of those
quanta. These other magnitudes are those which can be expressed as functions
of $c$ and $*$. They include the following:
\def\labelitemi{--}
\begin{itemize}\itemsep=1pt
\item Electrical resistance \ $(W_e)=c^{-1}=3.335600\ti10^{-11}\ {\rm
s}\cdot{\rm cm}^{-1}$.
\item Planck constant \ $h=\que{2\pi}\al\ *\cdot c$.
\item Elementary charge \ $e=(*)^{1/2}\cdot c$.
\item Flow of magnetic induction \ $(\va_e)=(*)^{1/2}$.
\item Magnetic field \ $(B_e)=(*)^{1/2}c^{-1}$.
\item Magnetic permeability \ $(\mu)=c^{-2}$.
\item Momentum of rotation \ $\hbar=\que *\al c$.
\end{itemize}

Where there is no symbol to designate the quantum, we have used that of the
magnitude in question, placing it in brackets.
\end{enumerate}

\section*{REFERENCES}

[1] R. Alvargonz\'alez and L. S. Soto:
``Zero-point radiation and the Big-Bang".\ {\em arXiv}: 0705 3722 VI [physics
gen-phis], 25 May 2007.

[2] R. Alvargonz\'alez and L. S. Soto:
``An analysis of the Big-Bang theory according to classical
physics". {\em arXiv-physics}/0408016.VI, 3 Aug 2004.

[3] R. Alvargonz\'alez:
{\em Interactions between zero-point radiation and electrons.}
{\em arXiv-physics/0311139}

[4] R. Alvargonz\'alez:
``On the magnitude of the energy flow inherent in zero-point radiation". {\em
arXiv: physics}/0311027v2 [physics gen-ph] 17 Feb 2004.

[5] R. Alvargonz\'alez and L. S. Soto:
``Zero-point radiation inertia and gravitation".\
{\em arXiv-physics}/0312096v3 [physics gen-ph] 26 Jun 2007.

[6] R. Alvargonz\'alez and L. S. Soto:
``Suggestions on Photons and on Fermions".
{\em arXiv:} 0705.3549 vI [physics. gen-ph] 24 May 2007.

[6] M. J. Sparnaay:
{\em Physica 24}, {\bf 751} (1958).

[7] S. K. Lamoreaux:
{\em Phys. Rev. Let. 6}, {\bf 78, 5} (1997).

[8] H. Boyer Timothy:
{\em Phys. Rev.} {\bf 182}, p. 497 (1949).

\end{document}